\documentclass[journal,10pt,twocolumn]{IEEEtran}

\usepackage{etoolbox}
\newtoggle{doublecolumn}
\newtoggle{calculateFigures}
\toggletrue{doublecolumn}
\togglefalse{calculateFigures}
\usepackage{etex}

\usepackage[english]{babel}
\usepackage{lipsum}
\usepackage{amsbsy,amsmath,amsfonts,amssymb,amsthm}
\usepackage{mathtools}
\usepackage{textcomp} 
\usepackage{relsize}

\usepackage{bm,cite,cases,pstricks,times,url,verbatim} 
\usepackage[noend]{algpseudocode}
\usepackage{booktabs}
\usepackage{tabularx,array,dcolumn,multirow}

\usepackage{algorithm}

\usepackage{amsbsy}

\usepackage{yfonts}

\usepackage{dutchcal} 

\usepackage{xr} 

\usepackage{soul} 

\usepackage{blkarray,bigdelim} 

\allowdisplaybreaks[4]

\usepackage{centernot} 

\newcommand{\subparagraph}{}

\iftoggle{doublecolumn}{
    \newtheorem{thm}{Theorem}
    \newtheorem{fact}{Fact}
    \newtheorem{lemma}{Lemma}
    \newtheorem{definition}{Definition}
    \newtheorem{conj}{Conjecture}
    \newtheorem{propos}{Proposition}
    \newtheorem{corol}{Corollary}
    \newtheorem{ass}{Assumption}
    \newtheorem{example}{Example}
    \newtheorem{remark}{Remark}
    \newtheorem{note}{Note}
    \newtheorem{obs}{Observation}
}{

    \newtheoremstyle{exampstyle}
      {0} 
      {0} 
      {\itshape} 
      {} 
      {\bfseries} 
      {.} 
      {.5em} 
      {} 

    \theoremstyle{exampstyle} \newtheorem{thm}{Theorem}
    \theoremstyle{exampstyle} 
    \theoremstyle{exampstyle} 
    \theoremstyle{exampstyle} 
    \theoremstyle{exampstyle} 
    \theoremstyle{exampstyle} 
    \theoremstyle{exampstyle} 
    \theoremstyle{exampstyle} 
    \theoremstyle{exampstyle} 
    \theoremstyle{exampstyle} 
    \theoremstyle{exampstyle} 
    \theoremstyle{exampstyle} 
}

\makeatletter
\newcommand{\pushright}[1]{\ifmeasuring@#1\else\omit\hfill$\displaystyle#1$\fi\ignorespaces}
\newcommand{\pushleft}[1]{\ifmeasuring@#1\else\omit$\displaystyle#1$\hfill\fi\ignorespaces}

\begingroup
\catcode`\#=11
\gdef\noautorotate{-dAutoRotatePages#/None}
\endgroup

\usepackage{caption}
\usepackage{subcaption}
\usepackage{graphicx}
\usepackage{epsf,bm}
\usepackage{epstopdf}

\newcommand{\subalign}[1]{%
  \vcenter{%
    \Let@ \restore@math@cr \default@tag
    \baselineskip\fontdimen10 \scriptfont\tw@
    \advance\baselineskip\fontdimen12 \scriptfont\tw@
    \lineskip\thr@@\fontdimen8 \scriptfont\thr@@
    \lineskiplimit\lineskip
    \ialign{\hfil$\m@th\textstyle##$&$\m@th\textstyle{}##$\crcr
      #1\crcr
    }%
  }
}

\iftoggle{calculateFigures}{
    \usepackage[crop=off,runs=2,pspdf=\noautorotate]{auto-pst-pdf}
}{
    \usepackage[off]{auto-pst-pdf}
}

\usepackage[unicode]{hyperref}
\newcommand\defeq{\stackrel{\mathclap{{\rm def}}}{=}}

\graphicspath{%
{./}
}
\def\BSTATE{\STATE\hskip-\ALG@thistlm}
\makeatother

\usepackage{color}

\usepackage{caption}
\usepackage{subcaption}
\begin{document}
    
    \author{\IEEEauthorblockN{Mohamed~A.~Abd-Elmagid\IEEEauthorrefmark{1}, Alessandro~Biason\IEEEauthorrefmark{2}, Tamer~ElBatt\IEEEauthorrefmark{1}\IEEEauthorrefmark{3}, Karim G. Seddik\IEEEauthorrefmark{4} and~Michele~Zorzi\IEEEauthorrefmark{2}}\\
        \IEEEauthorblockA{\small \IEEEauthorrefmark{1} Wireless Intelligent Networks Center (WINC), Nile University, Giza, Egypt}\\
        \IEEEauthorblockA{\small \IEEEauthorrefmark{2} Department of Information Engineering, University of Padova - via Gradenigo
            6b, 35131 Padova, Italy}\\
        \IEEEauthorblockA{\small \IEEEauthorrefmark{3} Dept. of EECE, Faculty of Engineering, Cairo University, Giza, Egypt}\\
        \IEEEauthorblockA{\small \IEEEauthorrefmark{4} Electronics and Communications Engineering Department, American University in Cairo, AUC Avenue, New Cairo 11835, Egypt}\\
        \IEEEauthorblockA{\small email: m.abdelaziz@nu.edu.eg, biasonal@dei.unipd.it, telbatt@ieee.org, kseddik@aucegypt.edu, zorzi@dei.unipd.it}
    }

    \title{Non-Orthogonal Multiple Access Schemes in Wireless Powered Communication Networks}
    \maketitle
    \pagestyle{empty}
    \thispagestyle{empty}

    \fontdimen2\font=3.3pt
    \begin{abstract}
        We characterize time and power allocations to optimize the sum-throughput of a Wireless Powered Communication Network (WPCN) with Non-Orthogonal Multiple Access (NOMA). In our setup, an Energy Rich (ER) source broadcasts wireless energy to several devices, which use it to simultaneously transmit data to an Access Point (AP) on the uplink. Differently from most prior works, in this paper we consider a generic scenario, in which the ER and AP do not coincide, i.e., are two separate entities. We study two NOMA decoding schemes, namely Low Complexity Decoding (LCD) and Successive Interference Cancellation Decoding (SICD). For each scheme, we formulate a sum-throughput optimization problem over a finite horizon. Despite the complexity of the LCD optimization problem, due to its non-convexity, we recast it into a series of geometric programs. On the other hand, we establish the convexity of the SICD optimization problem and propose an algorithm to find its optimal solution. Our numerical results demonstrate the importance of using successive interference cancellation in WPCNs with NOMA, and show how the energy should be distributed as a function of the system parameters.
    \end{abstract}
    
    \section{Introduction}

    In the past few years, there has been rapidly growing interest in developing new strategies and technologies for prolonging the lifetime of mobile devices. Among a variety of technologies, Energy Transfer (ET) has recently emerged as a promising solution. Although ET can be considered as a new resource for the mobile nodes, it also adds a layer of complexity to the system design and optimization. The goal of this paper is to study a particular application of ET, and derive the optimal strategies to maximize the network throughput.
    
    The concept of energy transfer has been analyzed in different fields. For example, in the \emph{energy cooperation} paradigm~\cite{5,6,7}, different devices transfer energy among themselves to equalize the energy level of the network and improve the overall performance. In addition, with the Simultaneous Wireless Information and Power Transfer (SWIPT) schemes, it became possible to send a signal carrying both information and energy simultaneously~\cite{1,2,3,4}. SWIPT was first studied from an information-theoretic perspective in~\cite{1,2}. The fundamental trade-off between simultaneously transmitting information and harvesting energy is studied for narrowband noisy channels in~\cite{1} and for frequency-selective channels in~\cite{2}. Afterwards, from a communication-theoretic perspective, \cite{3}~characterized the fundamental trade-off between transmitting energy and transmitting information over a point-to-point noisy link. Since modern energy harvesting circuits are unable to harvest energy and decode information simultaneously, \cite{4}~proposed two practical receiver designs, namely, time switching and power splitting. Under the time switching setting, the receiving antenna periodically switches between energy reception and information decoding phases. Instead, for the power splitting scheme, the received signal is split into two streams with different power levels; one is sent to the energy harvesting circuitry and the other to the information decoder. 
    
    Wireless Powered Communication Networks (WPCNs), a newly emerging type of wireless networks, has recently attracted considerable attention in the literature \cite{8,9,10,11,12,13,14}. In a WPCN, the devices first harvest wireless energy from a dedicated Energy Rich (ER) source, and then use it to uplink data to the Access Point (AP). \cite{8} assumed that ER and AP coincide and characterized the optimal time allocations to achieve the maximum sum-throughput and the max-min throughput. Although in a large body of the literature the ER and AP coincide, in this paper we consider them as separate entities to accommodate a more general setting. \cite{9}~exploited a data-cooperation technique to address the doubly near-far phenomenon that leads to unfair rate allocation among different users, as observed in~\cite{8}. However, this technique is suitable only for a smaller set of scenarios in which the terminal devices are closely placed. Moreover, it leads to a higher computational complexity to derive the scheduling policy. \cite{10}~studied a WPCN with heterogeneous nodes (nodes with and without RF energy harvesting capabilities) and showed how the presence of non-harvesting nodes can be utilized to enhance the network performance, compared to pure WPCNs~\cite{8}. Unlike prior slot-oriented optimization frameworks~\cite{8,9,10}, in which all the harvested energy is used in the same slot in which it is harvested, \cite{11}~focused on long-term optimization. 
    Although this incurs higher computational overhead, it also significantly improves the throughput of the network while maintaining fairness. Several papers investigated solutions in contrast to the limited harvested energy in WPCNs by introducing new paradigms~\cite{12,13,14}. 
    \cite{12}~extended the long-term maximization of the half-duplex case~\cite{11} to a full-duplex scenario, in which the ER and AP coincide and are able to broadcast wireless energy signals over the downlink and receive data signals over the uplink simultaneously. It was shown that the throughput region of the full-duplex scenario is notably larger than that of the half-duplex case~\cite{11}.
    \cite{13}~generalized conventional TDMA wireless networks (no energy harvesting) to a new type of wireless networks named generalized-WPCNs (g-WPCNs), where nodes are assumed to be equipped with RF energy harvesting circuitries along with energy supplies. It was shown that both conventional TDMA wireless networks and WPCNs with only RF energy harvesting nodes provide lower bounds on the performance of g-WPCNs in terms of maximum sum-throughput and max-min throughput.
    Although most of the literature has focused on orthogonal multiple access schemes (typically TDMA) for the uplink phase~\cite{8,9,10,11,12,13}, the authors in~\cite{14} introduced NOMA in WPCNs to enhance the power-bandwidth efficiency. Indeed, it was shown that NOMA improves spectral efficiency~\cite{15} with respect to orthogonal multiple access schemes. To work properly, NOMA calls for tuning the transmit power of the devices, so as to exploit interference cancellation techniques at the receiver side. However, \cite{14} focused on optimizing the time allocations to maximize the sum-throughput of the slot-oriented case (all the harvested energy in a slot is also consumed in the same slot). This approach, in turn, leads to sub-optimal polices since: 1) it does not optimize the transmit powers and 2) it does not take into account the global performance over a larger time horizon. Therefore, unlike~\cite{14}, in this paper we jointly optimize time and power allocations to maximize the sum-throughput over a finite horizon of $T$ slots. In our schemes, simultaneous transmissions can still be successful if the received signal power is sufficiently high, thus no fine synchronization is required. 
    
    Our contributions can be summarized as follows. 
    We propose Low Complexity Decoding and Successive Interference Cancellation Decoding, two schemes that aim at optimizing the sum-throughput of a WPCN with and without interference cancellation. Since LCD leads to a non-convex problem, we solve a sub-problem via casting it as a series of geometric programs. On the contrary, we formally establish the convexity of the sum throughput maximization problem with SICD and propose an algorithm to find the optimal transmission durations and powers. Our numerical results show the superiority of the interference cancellation scheme over the simpler LCD.
    
    \emph{Notation and Structure.} Subscripts ``$_{i,t}$'' denote the $i$-th node in the $t$-th time slot. Boldface letters are used to indicate all the elements of a quantity (e.g., $\mathbf{E} \defeq [E_{1,1},\ldots,E_{1,t},\ldots,E_{K,T}]$ or $\boldsymbol{\tau}_0 \defeq [\tau_{0,1},\ldots,\tau_{0,T}]$). With ``$\forall i$'' and ``$\forall t$'', we summarize $i = 1,\ldots,K$ and $t = 1,\ldots,T$, respectively.
    
    The paper is organized as follows. Section~\ref{sec:sys} presents the system model. Sections~\ref{sec:first} and~\ref{sec:SIC_based} present the LCD and SICD schemes, respectively. The numerical results are shown in Section~\ref{sec:num}. Finally, Section~\ref{sec:con} concludes the paper.
    
    \section{System Model} \label{sec:sys}
    
    \begin{figure}[!t]
        \centering
        \includegraphics[trim = 0mm 0mm 0mm 0mm, width=1\columnwidth]{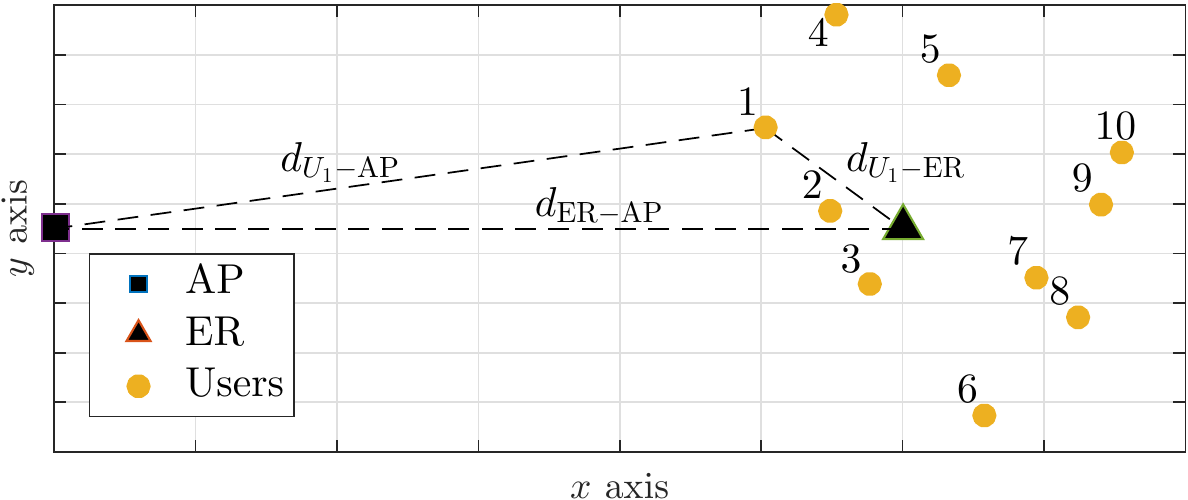}
        \caption{System model.}
        \label{fig:model}
    \end{figure}
    
    We study a WPCN composed of one AP, one ER source and $K$ users. The energy rich node is equipped with a stable energy supply and transfers wireless energy to the $K$ users in the network. User $U_{i}$, $i=1,\ldots, K$, receives the energy transferred by ER and uses the accumulated energy to send uplink data messages to AP. There are no other energy sources (neither environmental nor man-made) in the network.
    
    ER and all users are equipped with a single antenna each, operate over the same frequency band and the radios are half-duplex. Time is slotted and, without loss of generality, we assume that the slot duration is normalized to one. Every slot $t = 1,\ldots,T$ is divided in two phases. In the first $\tau_{0,t}$ seconds, ER broadcasts wireless energy on the downlink to recharge the batteries of the devices. In the remaining $1 - \tau_{0,t}$ seconds, all users transmit data to the AP independently and simultaneously.
    
    The positions of the users are known a priori, and thus their average channel gains are pre-estimated and known (in Section~\ref{sec:num} we will further discuss the channel models). The downlink channel power gain from ER to $U_{i}$ and the uplink channel power gain from $U_{i}$ to AP, during time slot $t$, are denoted by $h_{i,t}$ and $g_{i,t}$, respectively. Hence, the harvested energy by $U_{i}$ in the downlink phase is~
    \begin{align}\label{eq1}
        &\gamma_{i,t} \tau_{0,t} \defeq \eta_{i} h_{i,t} P_{B} \tau_{0,t}.
    \end{align}
    
    \noindent where $\eta_i$ denotes the efficiency of the energy harvesting circuitry,\footnote{$\eta_i$ depends on the efficiency of the harvesting antenna, the impedance matching circuit and the voltage multipliers. For example, we will use $\eta_i = 0.49$ according to the specifications of the commercial product P2110~\cite{powercast}.} and $P_{B}$ is the average transmit power by ER within $\tau_{0,t}$.
    
    \section{Low Complexity Decoding Scheme}
    \label{sec:first}
    In this section, we present the Low Complexity Decoding (LCD) scheme, in which the AP uses single-user decoders to detect the users' signals without performing interference cancellation. In particular, each user may interfere with all the others. Hence, we can express the average Signal-to-Interference-plus-Noise Ratio (SINR) at the AP for $U_{i}$ in time slot $t$ as~
    \begin{align}\label{eq:SINR}
        x_{i,t} \defeq \dfrac{g_{i,t} E_{i,t}}{\sigma^{2} \left(1 - \tau_{0,t}\right) + \sum_{\substack{j=1 \\ j \neq i}}^{K}{g_{j,t} E_{j,t}}},
    \end{align}
    
    \noindent where $E_{i,t}$ denotes the amount of consumed energy by $U_{i}$ in time slot $t$ and $\sigma^{2}$ denotes the noise power at the AP. According to Shannon's formula, and with the Gaussian approximation for interference, the achievable throughput of $U_{i}$, in time slot $t$, is given by~
    \begin{align}\label{eq3}
        R_{i,t} \defeq \left(1 - \tau_{0,t}\right)\log_{2}\left(1 + x_{i,t}\right).
    \end{align}
    
    Our objective is to characterize the maximum achievable sum-throughput over a finite horizon of $T$ time slots subject to energy causality constraints and practical decoding constraints. The energy causality constraints guarantee that, in slot $t$, only the energy harvested in slots $\leq t$ can be used. For the $i$-th user, it can be expressed as follows~
    \begin{align}\label{eq4}
        \sum_{n=1}^{t}{E_{i,n}} \leq \sum_{n=1}^{t}{\gamma_{i,n} \tau_{0,n}}, \qquad \forall t.
    \end{align}
    
    \noindent  Without interference cancellation techniques (which will be discussed in Section~\ref{sec:SIC_based}), the SINR $x_{i,t}$ may turn out to be very low. If it falls under a pre-specified threshold value $S_i^{\rm th}$, we assume that decoding is not possible, i.e., we impose the following constraint~
    \begin{align}\label{eq5}
        x_{i,t} \geq S_{i}^{\rm th}, \quad\forall i, \forall t.
    \end{align}
    
    Using the previous expressions, we can now formulate the sum-throughput maximization problem as~
    \begin{subequations} \label{eq6}
        \begin{flalign}\label{eq:P1:objective}
            \text{\textbf{P}$_{\rm LCD}$:} && & \max_{\boldsymbol{\tau}_0, \mathbf{E}, \mathbf{x}}\sum_{t=1}^{T}\sum_{i=1}^{K} R_{i,t}, &
        \end{flalign}
        \vspace{-\belowdisplayskip}
        \vspace{-\abovedisplayskip}
        \begin{alignat}{3}
            \shortintertext{subject to}
            & \mbox{Eqs.~\eqref{eq:SINR}, \eqref{eq4}, \eqref{eq5}}, \quad && \forall i,\ &&\forall t, \\
            & 0 \leq \tau_{0,t} \leq 1, \quad && && \forall t, \\
            &  E_{i,t} \geq 0, \quad && \forall i,\ &&\forall t,
        \end{alignat}
    \end{subequations}
    \noindent where $\boldsymbol{\tau}_0$, $\mathbf{E}$ and $\mathbf{x}$ are vectors whose elements are the harvesting time durations, the energy consumed by each user and the average SINR at AP for each user over the finite horizon of $T$ slots, respectively. Note that variable $x_{i,t}$ can be substituted using~\eqref{eq:SINR} and removed from the problem. The objective function of \textbf{P}$_{\rm LCD}$ is not convex, thus \textbf{P}$_{\rm LCD}$ is a non-convex optimization problem. However, if $\boldsymbol{\tau}_0$ is given, \textbf{P}$_{\rm LCD}$ can be transformed into a series of separate Geometric Programs (GPs). Indeed, thanks to the monotonicity property of the $\log$ function, for a fixed $\boldsymbol{\tau}_0$, an equivalent optimization problem to \textbf{P}$_{\rm LCD}$ can be written as follows 
    \begin{subequations} \label{eq:P2}
        \begin{flalign}\label{eq:P2:objective}
            \text{\textbf{P}$_{\rm LCD}(\boldsymbol{\tau}_0)$:} && & \min_{\mathbf{E}, \mathbf{x}} \frac{1}{\prod_{t=1}^{T}\prod_{i=1}^{K}{\left(1 + x_{i,t}\right)^{1 - \tau_{0,t}}}}, &
        \end{flalign}
        \vspace{-\belowdisplayskip}
        \vspace{-\abovedisplayskip}
        \begin{alignat}{3}
            \shortintertext{subject to}
            \label{eq:P2a}& x_{i,t} \times \dfrac{\sigma^{2}(1 - \tau_{0,t}) + \sum_{\substack{j=1 \\ j \neq i}}^{K}{g_{j,t} E_{j,t}}}{g_{i,t} E_{i,t}}
            \leq 1, \quad && \forall i,\ &&\forall t \\
            \label{eq:P2b}& \dfrac{\sum_{n=1}^{t}{E_{i,n}}}{\sum_{n=1}^{t}{\gamma_{i,n} \tau_{0,n}}} \leq 1, \quad && \forall i,\ &&\forall t, \\
            \label{eq:P2c} & S^{\rm th}_{i} x_{i,t}^{-1} \leq 1, \quad && \forall i,\ &&\forall t \\
            & E_{i,t} \geq 0, \quad && \forall i,\ &&\forall t.
        \end{alignat}
    \end{subequations}
    
    It can be verified that constraints~\eqref{eq:P2a}-\eqref{eq:P2c} are expressed in the standard form for a Geometric Program (GP). On the other hand, the objective function of~\textbf{P}$_{\rm LCD}(\boldsymbol{\tau}_0)$ is a ratio between two posynomial functions, thus \textbf{P}$_{\rm LCD}(\boldsymbol{\tau}_0)$ is a non-convex complementary GP~\cite{17,18}. Since directly solving complementary GPs is NP-hard, we introduce an approximate approach, which allows us to solve \textbf{P}$_{\rm LCD}(\boldsymbol{\tau}_0)$ iteratively using standard GPs solvers. We will show that the approximate solution satisfies the Karush-Kuhn-Tucker (KKT) conditions of \textbf{P}$_{\rm LCD}(\boldsymbol{\tau}_0)$ and thus is guaranteed to be a local optimal solution for \textbf{P}$_{\rm LCD}(\boldsymbol{\tau}_0)$. 
    
    \subsection{Approximate Solution of \emph{\textbf{P}}$_{\rm LCD}(\texorpdfstring{\boldsymbol{\tau}_0}{})$ }
    
    We approximate the denominator of~\eqref{eq:P2:objective}, namely $f(\mathbf{x})$, with a monomial function $\tilde{f}(\mathbf{x})$. In this case, the new approximate optimization problem becomes a standard GP that can be solved iteratively using standard techniques~\cite{20}. In particular, if $\tilde{f}(\mathbf{x})$ satisfies the following three conditions~\cite{19}, then the solution of the series of approximate optimization problems converges to a point satisfying the KKT conditions of the original problem \textbf{P}$_{\rm LCD}(\boldsymbol{\tau}_0)$:~
    \begin{subequations}
        \label{eq:three_conditions}
        \begin{align}
            \label{eq:three_conditions:1} & 1) \ f(\mathbf{x}) \geq \tilde{f}(\mathbf{x}), \ \forall \mathbf{x}, \\
            \label{eq:three_conditions:2} & 2) \ \tilde{f}(\bar{\mathbf{x}}) = f(\bar{\mathbf{x}}), \\
            \label{eq:three_conditions:3} & 3) \ \nabla \tilde{f}(\bar{\mathbf{x}}) = \nabla f(\bar{\mathbf{x}}),
        \end{align}
    \end{subequations}
    
    
    \noindent where $\bar{\mathbf{x}}$ is the solution of the approximate GP in the previous iteration.
    
    Now, we introduce the approximate objective function that satisfies the conditions given in~\eqref{eq:three_conditions}. In particular, we choose~
    \begin{align}\label{eq9}
        \tilde{f}(\mathbf{x}) = c\prod_{t=1}^{T}\prod_{i=1}^{K}{\left(x_{i,t}\right)^{y_{i,t}(1 - \tau_{0,t})}}. 
    \end{align}
    
    %
    \noindent Equation~\eqref{eq:three_conditions:2} yields~
    \begin{align} \label{eq10}
        c = \dfrac{\prod_{t=1}^{T}\prod_{i=1}^{K}{\left(1 + \bar{x}_{i,t}\right)^{1 - \tau_{0,t}}}}{\prod_{t=1}^{T}\prod_{i=1}^{K}{\left(\bar{x}_{i,t}\right)^{y_{i,t}\left(1 - \tau_{0,t}\right)}}}.
    \end{align}
    
    \noindent Moreover, from~\eqref{eq:three_conditions:3}, we have~
    \begin{align}\label{eq11}
        y_{i,t} = \dfrac{\bar{x}_{i,t}}{1 + \bar{x}_{i,t}}, \qquad \forall i, \ \forall t.
    \end{align}

    \noindent Finally, by substituting $c$ and $y_{i,t}$ into $\tilde{f}(\mathbf{x})$, we can write the condition in~\eqref{eq:three_conditions:1} as~
    \begin{align}\label{eq35}
        \prod_{t,i}{(1 + x_{i,t})^{1 - \tau_{0,t}}} \geq \prod_{t,i}{(1 + \bar{x}_{i,t})^{1 - \tau_{0,t}} \left(\dfrac{x_{i,t}}{\bar{x}_{i,t}}\right)^{y_{i,t}\left(1 - \tau_{0,t}\right)}}.
    \end{align}
    
    \noindent We note that the previous inequality should hold for every $\mathbf{x}$. This is achieved if the following condition holds $\forall i$, $\forall t$,~
    \begin{align}\label{eq36}
        (1 + x_{i,t})^{1 - \tau_{0,t}} \geq (1 + \bar{x}_{i,t})^{1 - \tau_{0,t}} \left(\dfrac{x_{i,t}}{\bar{x}_{i,t}}\right)^{y_{i,t}\left(1 - \tau_{0,t}\right)},
    \end{align}
    
    \noindent which is equivalent to~
    \begin{align}\label{eq37}
        G(x_{i,t}) \defeq \log \Bigg(\dfrac{1 + \bar{x}_{i,t}}{1 + x_{i,t}}\times\left(\dfrac{x_{i,t}}{\bar{x}_{i,t}}\right)^{\dfrac{\bar{x}_{i,t}}{1 + \bar{x}_{i,t}}}\Bigg) \leq 0.
    \end{align}
    
    \noindent By taking the derivatives of $G(x_{i,t})$ over $x_{i,t}$, we obtain~
    \begin{align}\label{eq38}
        &\dfrac{\partial}{\partial x_{i,t}}G(x_{i,t}) = \dfrac{\bar{x}_{i,t}}{x_{i,t}\left(1 + \bar{x}_{i,t}\right)} - \dfrac{1}{1 + x_{i,t}}, \\
        &\dfrac{\partial^2}{\partial x_{i,t}^2}G(\bar{x}_{i,t}) = \dfrac{-1}{\bar{x}_{i,t}\left(1 + \bar{x}_{i,t}\right)^2}.\label{eq39}
    \end{align}
    
    \noindent From~\eqref{eq38} and~\eqref{eq39}, it follows that $G(x_{i,t})$ is a convex downward function for $x_{i,t} \geq 0$ and its maximum is attained at $x_{i,t} = \bar{x}_{i,t}$, with $G(\bar{x}_{i,t}) = 0$. Therefore, $G(x_{i,t}) \leq 0$ for $x_{i,t} \geq 0$ is satisfied and Condition~\eqref{eq:three_conditions:1} holds. Thus, since~\eqref{eq:three_conditions:2}-\eqref{eq:three_conditions:3} hold by constructions, all the conditions are satisfied and the solution of the approximate problem is a KKT point of~\textbf{P}$_{\rm LCD}(\boldsymbol{\tau}_0)$.
    
    We now present the steps of Algorithm~\ref{alg:P2}. Starting with an initial $\bar{\mathbf{x}}$, we can obtain  $c$ and $y_{i,t}$ from~\eqref{eq10} and~\eqref{eq11}, respectively. With these values, we solve the approximate geometric program. The obtained solution can be used to get new values of $c$ and $y_{i,t}$. The procedure is repeated until the sum-throughput converges to a predetermined accuracy.

    \begin{algorithm}[H]
        \caption{(\textbf{P}$_{\rm LCD}(\boldsymbol{\tau}_0)$ solver)}\label{alg:P2}
        \begin{algorithmic}[1]
            \State  Initialize $\bar{\mathbf{x}}$
            \State  Compute $c$ and $y_{i,t}$ using \eqref{eq10} and \eqref{eq11}
            \Repeat
            \State  Solve the approximate \textbf{P}$_{\rm LCD}(\boldsymbol{\tau}_0)$
            \State  Update $c$ and $y_{i,t}$ using \eqref{eq10} and \eqref{eq11}, respectively
            \Until{sum-throughput converges}
        \end{algorithmic}
    \end{algorithm}
    
    We finally note that $\boldsymbol{\tau}_0$ is \emph{given} in~\textbf{P}$_{\rm LCD}(\boldsymbol{\tau}_0)$, and finding its optimal value (i.e., solving the original problem~\textbf{P}$_{\rm LCD}$) is beyond the scope of this paper.
    
    \section{Successive Interference Cancellation Decoding Scheme}\label{sec:SIC_based}
    
    In the scheme described in Section~\ref{sec:first}, each user suffers interference, at the AP, from all other users in the network. Although this significantly reduces the decoding complexity, it leads to sharp degradation in the achievable maximum sum-throughput. Nevertheless, it is possible to use more sophisticated decoding techniques to partially solve the problem and further enhance the performance of the system. In particular, in this section we introduce a Successive Interference Cancellation Decoding~(SICD) scheme and characterize the associated maximum sum-throughput.
    
    With SICD, when the signal received by the AP for one user is decoded, it can be removed from the interference term of the other users, leading to better SINRs. In this paper, we adopt a fixed order decoding strategy and, without loss of generality, we follow the order of the indices (i.e., user $U_i$ suffers interference from all other users with indices $i+1,\ldots,K$).  Hence, the signal of $U_{K}$ is decoded without any interference. Part of our future work includes the study of dynamic decoding strategies, so as to prioritize the users according to their channels and improve the fairness.

    
    The SINR of $U_{i}$ in time slot $t$, after interference cancellation, can be expressed as~
    \begin{align}\label{eq12}
        x_{i,t} = \dfrac{g_{i,t} E_{i,t}}{\sigma^{2} \left(1 - \tau_{0,t}\right) + \sum_{j = i+1}^{K}{g_{j,t} E_{j,t}}}, \quad \forall i,
    \end{align}
    
    \noindent where we imposed $\sum_a^b (\cdot) = 0$ if $a > b$. The achievable throughput for $U_{i}$ can be expressed as in~\eqref{eq3} using~\eqref{eq12}. Hence, it can be shown that the achievable sum-throughput, in time slot $t$, is given by~
    \begin{align}\label{eq14}
        R_{\rm sum}^{(t)} \defeq \left(1 - \tau_{0,t}\right)\log_{2} \left(1 + \dfrac{\sum_{i=1}^{K}{g_{i,t} E_{i,t}}}{\sigma^{2}\left(1 - \tau_{0,t}\right)}\right).
    \end{align}
    
    \noindent The previous expression is significantly different from~\eqref{eq:P1:objective} due to the absence of the interference term from the denominator of the fraction inside the $\log$ function, and will allow us to formulate a \emph{convex} optimization problem, unlike in Section~\ref{sec:first}. We now present the analogous of \textbf{P}$_{\rm LCD}$ when SICD is taken into account:
    \begin{subequations} \label{eq:P3}
        \begin{flalign}\label{obj_p3}
            \text{\textbf{P}$_{\rm SICD}$:} && & \max_{\boldsymbol{\tau}_0, \mathbf{E}, \mathbf{x}}\sum_{t=1}^{T}\sum_{i=1}^{K} R_{i,t}, &
        \end{flalign}
        \vspace{-\belowdisplayskip}
        \vspace{-\abovedisplayskip}
        \begin{alignat}{2}
            \shortintertext{subject to}
            & \mbox{Eqs.~\eqref{eq4}, \eqref{eq5}, \eqref{eq12}}, \quad && \forall i, \forall t, \\
            & 0 \leq \tau_{0,t} \leq 1, \quad && \forall t, \\
            & E_{i,t} \geq 0, \quad && \forall i, \forall t.
        \end{alignat}
    \end{subequations}
    
    \begin{thm}\label{th:1} 
        \emph{\textbf{P}$_{\rm SICD}$} is a convex optimization problem.  
        \begin{IEEEproof}
            We first recall that if $f(x)$ is concave, then also its perspective function $g(x,t) = t f(x/t)$ is concave (see~\cite[Section~3.2.6]{20}). Note that, using $t = 1 - \tau_0^{(t)}$, the sum-throughput $R_{\rm sum}^{(t)}$ is the perspective function of the concave function $\log_{2} \left(1 + \dfrac{\sum_{i=1}^{K}{g_{i}^{(t)} E_{i}^{(t)}}}{\sigma^{2}}\right)$. Therefore, $R_{\rm sum}^{(t)}$ is a concave function in $[\tau_0^{(t)},E_{1}^{(t)},\ldots,E_{K}^{(t)}]$. Since a non-negative weighted sum of concave functions is also concave, then the objective function of \emph{\textbf{P}$_{\rm SICD}$} in \eqref{obj_p3} which is the non-negative weighted summation of $R_{\rm sum}^{(t)}$, $\forall t,$ is a concave function in $(\boldsymbol{\tau}_{0},\mathbf{E})$. In addition, all constraints of \emph{\textbf{P}$_{\rm SICD}$} are affine in $(\boldsymbol{\tau_0},\mathbf{E})$ and the proof is complete.
        \end{IEEEproof}  
    \end{thm}

    Based on Theorem \ref{th:1}, \textbf{P}$_{\rm SICD}$ is a convex optimization problem, and hence can be solved using standard convex optimization tools. The Lagrangian of \textbf{P}$_{\rm SICD}$ is given by~
    \begin{align}\label{eq16}
        \mathcal{L}\left(\mathbf{E}, \boldsymbol{\tau_0}, \boldsymbol{\lambda}, \boldsymbol{\mu}\right) = &\ \sum_{t=1}^{T}\sum_{i=1}^{K} R_{i,t} \nonumber\\ 
        & + \sum_{i=1}^{K}\sum_{n=1}^{T} {\lambda_{i,n} \left(\sum_{t=1}^{n}{\left(\gamma_{i,t} \tau_{0,t} - E_{i,t}\right)}\right)} \nonumber\\
        & + \sum_{i=1}^{K} \sum_{t=1}^{T}{\mu_{i,t}\left(x_{i,t} - S_{i}^{\rm th}\right)},
    \end{align}
    
    \noindent where $\lambda_{i,t}$ and $\mu_{i,t}$ are the dual variables associated with constraints~\eqref{eq4} and~\eqref{eq5}, respectively. Hence, the dual function, denoted by $\mathcal{G}\left(\boldsymbol{\lambda},\boldsymbol{\mu}\right)$, is obtained by solving the following optimization problem
    \begin{subequations} \label{eq:P4}
        \begin{flalign}
            \text{\textbf{D}$_{\rm SICD}$:} && & \max_{\boldsymbol{\tau}_0, \mathbf{E}} \mathcal{L}\left(\mathbf{E}, \boldsymbol{\tau_0}, \boldsymbol{\lambda}, \boldsymbol{\mu}\right), &
        \end{flalign}
        \vspace{-\belowdisplayskip}
        \vspace{-\abovedisplayskip}
        \begin{alignat}{2}
            \shortintertext{subject to}
            & 0 \leq \tau_{0,t} \leq 1, \quad && \forall t, \\
            & E_{i,t} \geq 0, \quad && \forall i, \forall t.
        \end{alignat}
    \end{subequations}
    
    \noindent Consequently, the dual problem will be: $\underset{\boldsymbol{\lambda}, \boldsymbol{\mu} \geq 0}{\text{min}} \; \mathcal{G}\left(\boldsymbol{\lambda}, \boldsymbol{\mu}\right)$. We now propose an algorithm to solve it.
    
    \begin{thm}\label{th:2} 
        Given $\boldsymbol{\lambda}$ and $\boldsymbol{\mu}$, the optimal time and energy allocations of \emph{\textbf{D}$_{\rm SICD}$} are given by~
        \begin{align}
            &\hspace{-.2cm}\tau_{0,t}^\star = \min\left[\left(1 - \dfrac{\sum_{i=1}^{K}{g_{i,t} E_{i,t}}}{z_t^\star\sigma^{2}}\right)^+, \;\; 1\right],\label{eq18} \\
            &\hspace{-.2cm}E_{i,t}^\star =  \Bigg(\dfrac{(1 - \tau_{0,t})(g_{i,t} - \sigma^2 a_{i,t})}{a_{i,t}g_{i,t}} - \dfrac{1}{g_{i,t}}\sum_{\substack{j=1 \\ j \neq i}}^{K}{g_{j,t} E_{j,t}}\Bigg)^+ \!\!\!. \label{eq19}
        \end{align}
        
        \noindent $a_{i,t}$ is defined as~
        \begin{align}\label{eq20}
            \begin{split}
                a_{i,t} \defeq \ln(2) \Bigg(&\sum_{n=t}^{T}\lambda_{i,n} + g_{i,t} \chi\{i \geq 2\}\sum_{j=1}^{i-1}{\mu_{j,t} S_{j}^{\rm th}} \\ 
                &-\mu_{i,t} g_{i,t} \Bigg),
            \end{split}
        \end{align}
        
        \noindent where $\chi\{\cdot\}$ is the indicator function and $(\cdot)^+ \defeq \max\{0,\cdot\}$. In addition, $z_t^\star$ is the unique solution of $f(z_t) = b^{(t)}$, where $f(z)$ and $b^{(t)}$ are given, respectively, by~
        \begin{align}
            &f(z_t) = \ln\left(1 + z_t\right) - \dfrac{z_t}{1 + z_t}, \label{eq21} \\
            &b^{(t)} = \ln(2) \left(\sigma^{2} \sum_{i=1}^{K}{\mu_{i,t}S_{i}^{\rm th}} + \sum_{i=1}^{K}\sum_{n=t}^{T}{\lambda_{i,n} \gamma_{i,t} }\right). \label{eq22}
        \end{align}
        
        \begin{IEEEproof}
            It can be easily shown that there exist $\boldsymbol{\tau_0}$ and $\mathbf{E}$ that strictly satisfy all the constraints of \emph{\textbf{D}$_{\rm SICD}$}. Hence, according to Slater's condition~\cite{20}, strong duality holds for this problem; therefore, the KKT conditions given below are necessary and sufficient for global optimality:~
            \begin{align}
                \dfrac{\partial}{\partial \tau_{0,t}} \mathcal{L} =&\ \ln\left(1 + \dfrac{\sum_{j=1}^{K}{g_{j,t} E_{j,t}}}{\sigma^{2}(1 - \tau_{0,t})}\right) \label{eq23}\\
                &\ - \dfrac{\sum_{j=1}^{K}{g_{j,t} E_{j,t}}}{\sigma^{2}(1 - \tau_{0,t}) + \sum_{j=1}^{K}{g_{j,t} E_{j,t}}} - b^{(t)} = 0, \nonumber\\
                \dfrac{\partial}{\partial E_{i,t}} \mathcal{L} =&\ \dfrac{\dfrac{g_{i,t}}{\sigma^{2} }}{1 + \dfrac{\sum_{i=1}^{K}{g_{i,t} E_{i,t}}}{\sigma^{2}(1 - \tau_{0,t})}} - a_{i,t} = 0,\label{eq24}
            \end{align}
            
            \noindent for every $i$ and $t$, where $a_{i,t}$ and $b^{(t)}$ are given by (\ref{eq20}) and (\ref{eq22}), respectively. By defining the variable $z_t = \dfrac{\sum_{j=1}^{K}{g_{j,t} E_{j,t}}}{\sigma^{2}\left(1 - \tau_{0,t}\right)}$, \eqref{eq23}~can be reformulated as $f(z_t) = b^{(t)}$, where $f(z_t)$ is given in (\ref{eq21}). It can be easily shown that $f(z_t)$ is a monotonically increasing function of $z_t \geq 0$, where $f(0) = 0$. Therefore, there exists a unique solution $z_t^\star$ that satisfies $f(z_t^\star) = b^{(t)}$ and, hence, $\tau_{0,t}^\star$ can be expressed as in~\eqref{eq18}. Finally, using~\eqref{eq24}, we obtain $E_{i,t}^\star$ as in~\eqref{eq19}, which concludes the proof.
        \end{IEEEproof}
    \end{thm}

    We now summarize how to solve \textbf{P}$_{\rm SICD}$ using Algorithm~\ref{alg:P3}. For a fixed $\boldsymbol{\lambda}$ and $\boldsymbol{\mu}$, we derive the optimal time and energy allocations using Theorem~\ref{th:2} through applying the alternating optimization procedure\footnote{Applying the alternating optimization procedure is guaranteed to converge to the global optimum since the Lagrangian of \textbf{P}$_{\rm SICD}$ is a concave function of $\boldsymbol{\tau}_{0}$ and $\mathbf{E}$ and a smooth function in both $\boldsymbol{\tau}_{0}$ and $\mathbf{E}$.} over the time slots. Afterwards, we update $\boldsymbol{\lambda}$ and $\boldsymbol{\mu}$ using the sub-gradient method with the sub-gradient of $G(\boldsymbol{\lambda},\boldsymbol{\mu})$ given by $[\nu_{i,n}\; \psi_{i,t}]$, where
    \begin{alignat}{2}
        &\nu_{i,n} = \sum_{n=1}^{t}{\left(\gamma_{i,t} \tau_{0,t}^\star - E_{i,t}^\star\right)}, \quad &&\forall i, \ \forall t, \label{eq33}\\
        &\psi_{i,t} = x_{i,t}^\star - S_i^{\rm th}, \quad &&\forall i, \ \forall t,\label{eq34}
    \end{alignat}
    then we use the updated dual variables to obtain the optimal time and energy allocations from Theorem \ref{th:2} again and so on until the stopping criteria of the sub-gradient method are met. Hence, the last updated dual variables will be the optimal solution of the dual problem, and the corresponding time and energy allocations, given by Theorem \ref{th:2}, will be the optimal solution of \textbf{P}$_{\rm SICD}$.

    \begin{algorithm}[ht]
        \caption{(\textbf{P}$_{\rm SICD}$ solver)}\label{alg:P3}
        \begin{algorithmic}[1]
            \State  Initialize $\boldsymbol{\lambda}$ and $\boldsymbol{\mu}$
            \Repeat
            \State  Initialize $\boldsymbol{\tau_0}$ and $\mathbf{E}$
            \Repeat
            \State  Update $\boldsymbol{\tau_0}$ and $\mathbf{E}$ using \eqref{eq18} and \eqref{eq19}
            \Until $\boldsymbol{\tau_0}$ and $\mathbf{E}$ converge
            \State  Update $\boldsymbol{\lambda}$ and $\boldsymbol{\mu}$ using the sub-gradient method
            \Until $\boldsymbol{\lambda}$ and $\boldsymbol{\mu}$ converge
            \State  Set $\boldsymbol{\tau_0^\star} = \boldsymbol{\tau_0}$ and $\mathbf{E^\star} = \mathbf{E}$
        \end{algorithmic}
    \end{algorithm}
    
    \begin{figure}[!t]
        \centering
        \includegraphics[trim = 0mm 0mm 0mm 0mm, width=1\columnwidth]{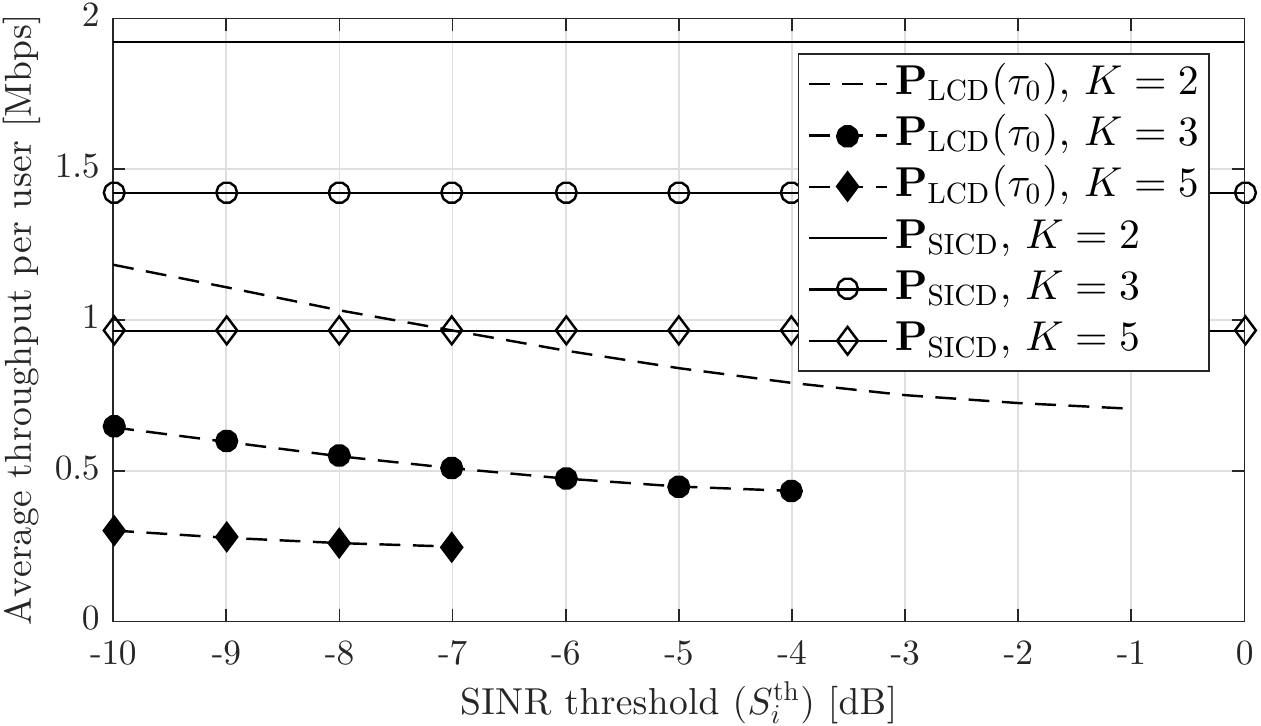}
        \caption{Average throughput per user vs. decodability threshold when $d_{\rm ER-AP} = 100$~m.}
        \label{fig:P1_P3}
    \end{figure}

    \section{Numerical results}
    \label{sec:num}
    
    We define $d_{\rm ER-AP}$ as the distance between ER and AP, the users are uniformly placed in a circle around ER at a distance $d_{U_i-{\rm ER}}$, and $d_{U_i-{\rm AP}}$ is the distance between $U_i$ and AP (see \figurename~\ref{fig:model}). 
    For the uplink transmission, the noise power is $-155$~dBm/Hz, the bandwidth is $1$~MHz and the path loss model is $g_{i,t} = 10^{-3} d_{U_i-{\rm AP}}^{-2}$. For the downlink we used the parameters of the P2110 device~\cite{powercast}, and in particular a fixed transmission power $P_B = 3$~W, a central frequency of $915$~MHz, a receiver antenna gain $G_r = 6$~dB, an efficiency $\eta_i = 0.49$ and Friis' formula for the downlink channel power gain derived with $d_{i,{\rm ER}} = 5$~m. Note that, as mentioned before, the slot duration is normalized to one.
    
    First, in \figurename~\ref{fig:P1_P3}, we compare \textbf{P}$_{\rm LCD}(\boldsymbol{\tau}_0)$ and \textbf{P}$_{\rm SICD}$ in terms of average throughput per user. 
    The geometric programs were solved using CVX, a package for specifying and solving convex programs~\cite{cvx}. Due to the numerical complexity of finding the optimal solution of \textbf{P}$_{\rm LCD}$, we use its simplified version \textbf{P}$_{\rm LCD}(\boldsymbol{\tau}_0)$ (defined in~\eqref{eq:P2}). In this case, we do not explicitly optimize $\boldsymbol{\tau}_0$, but we derive it from the solution of \textbf{P}$_{\rm SICD}$. Indeed, even if we managed to solve its simpler version \textbf{P}$_{\rm LCD}(\boldsymbol{\tau}_0)$, finding the optimal $\boldsymbol{\tau}_0$ for \textbf{P}$_{\rm LCD}$ would still be an open problem and is part of our future work. In \figurename~\ref{fig:P1_P3}, it can be seen that \textbf{P}$_{\rm LCD}(\boldsymbol{\tau}_0)$ has a solution only for very low values of the decodability threshold. Also, note that the greater the number of users, the sooner \textbf{P}$_{\rm LCD}(\boldsymbol{\tau}_0)$ becomes infeasible. Instead, the throughput of \textbf{P}$_{\rm SICD}$ is not influenced by $S_i^{\rm th}$, since, with SICD, the average SINR of every user is always greater than $0$~dB.

\begin{figure}[!t]
        \centering
        \includegraphics[trim = 0mm 0mm 0mm 0mm, width=1\columnwidth]{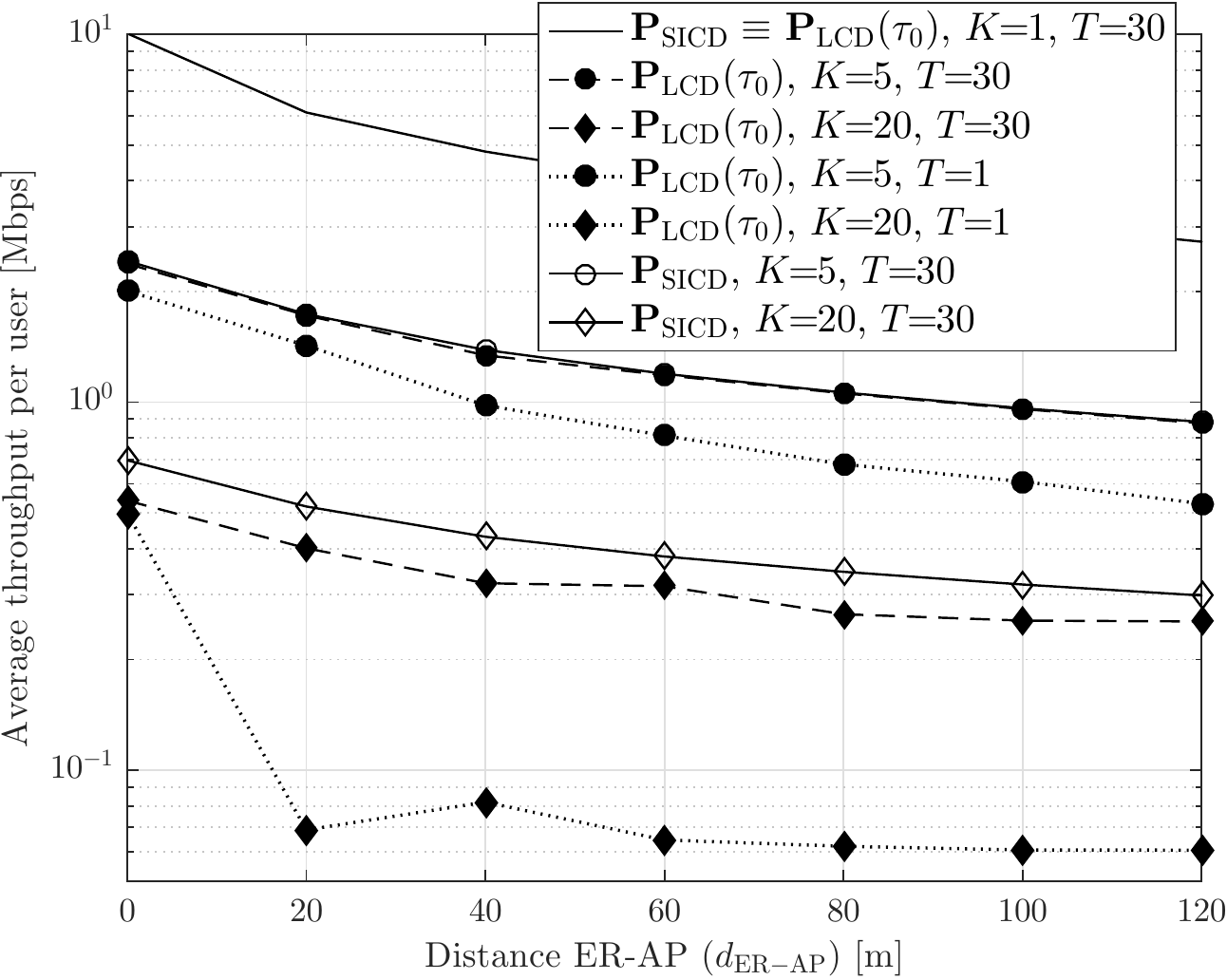}
        \caption{Average throughput per user (log-scale) of \textbf{P}$_{\rm SICD}$ vs. distance between ER and AP when $S_i^{\rm th} = 0$. The curves of \textbf{P}$_{\rm SICD}$ for $T = 1$ are not shown because they almost coincide with the case \textbf{P}$_{\rm SICD}$ and $T = 30$.}
        \label{fig:change_K}
    \end{figure}

\begin{figure}[!t]
        \centering
        \includegraphics[trim = 0mm 0mm 0mm 0mm, width=1\columnwidth]{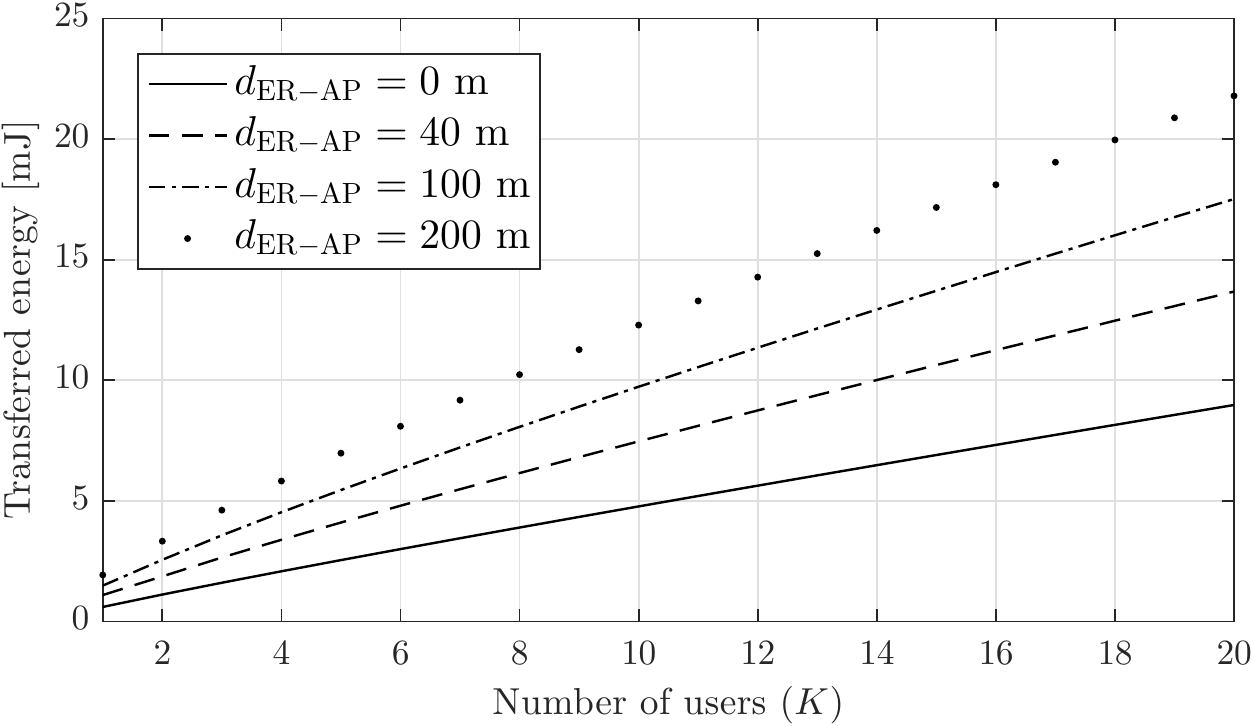}
        \caption{Downlink transferred energy of \textbf{P}$_{\rm SICD}$ to all users in all slots vs. number of users when $S_i^{\rm th} = 0$~dB.}
        \label{fig:plot_ET}
    \end{figure}

    In \figurename~\ref{fig:change_K}, we change the number of users in the system and derive the average throughput per user of \textbf{P}$_{\rm LCD}(\boldsymbol{\tau}_0)$ and \textbf{P}$_{\rm SICD}$ as a function of the distance $d_{\rm ER-AP}$. When only one user is considered, \textbf{P}$_{\rm LCD}(\boldsymbol{\tau}_0)$ and \textbf{P}$_{\rm SICD}$ coincide, whereas for $K > 1$ the scheme with interference cancellation always obtains better performance. It is worth noting that when the number of slots $T$ is large ($T = 30$), the solution of \textbf{P}$_{\rm LCD}(\boldsymbol{\tau}_0)$ closely approaches the solution of \textbf{P}$_{\rm SICD}$. However, when $T = 1$, the two schemes differ significantly. In practice, it can be verified that, since no minimum decodability threshold is imposed in this case, \textbf{P}$_{\rm LCD}(\boldsymbol{\tau}_0)$ degenerates to a pure TDMA scheme in which only one user at a time accesses the channel. This is not the case for SICD, in which all users transmit in every slot.
    Also, it can be seen that the greater $K$, the lower the average throughput per user. However, by multiplying every curve by the corresponding $K$, it can be verified that the sum-throughput increases with the number of users. For example, at $d_{\rm ER-AP} = 100$~m, \textbf{P}$_{\rm SICD}$ would obtain a sum-throughput of $3$~Mbps and $6.4$~Mbps for $K = 1$ and $K = 20$, respectively. This happens thanks to the broadcast nature of the energy transfer from ER. Indeed, until a certain physical threshold is reached, increasing the users corresponds to increasing the energy used in the network and thus the amount of data sent (provided that the SINR decodability constraints are met). This effect can also be seen in \figurename~\ref{fig:plot_ET}, where we plot the transferred energy over the downlink of \textbf{P}$_{\rm SICD}$ vs. the number of users, $K$. The total transferred energy is an increasing function $K$, because more terminals are able to receive the broadcast signal, and of the distance $d_{\rm ER-AP}$. Indeed, since in our scenario the users are located close to ER, the larger the distance between ER and AP, the stronger the path loss in uplink, thus more energy is required to compensate it.
    
    \section{Conclusion}
    We studied a wireless powered communication network with non-orthogonal multiple channel access. In our model, one energy rich source transfers energy to a group of nodes which utilize the received energy over the uplink to transmit to an access point. With the goal of maximizing the sum-throughput of the system, we proposed and solved two different decoding schemes for the data uplink phase, with and without successive interference cancellation at the receiver side. In particular, we jointly optimized the time and power allocations over a finite horizon. Our numerical results showed the great improvement in the achievable maximum sum-throughput, when employing successive interference cancellation schemes. They also demonstrated the impact of increasing the number of users, and the distance between the access point and the energy rich source, on the network performance.
    
    Part of the future work includes the study of online approaches to the sum-throughput problem, sub-optimal methods to find the downlink durations in the sum-throughput maximization problem with LCD and the comparison with classic orthogonal MAC schemes.
    \label{sec:con}
    
    \bibliography{mypaper}

\begin{thebibliography}{10}
\providecommand{\url}[1]{#1}
\csname url@samestyle\endcsname
\providecommand{\newblock}{\relax}
\providecommand{\bibinfo}[2]{#2}
\providecommand{\BIBentrySTDinterwordspacing}{\spaceskip=0pt\relax}
\providecommand{\BIBentryALTinterwordstretchfactor}{4}
\providecommand{\BIBentryALTinterwordspacing}{\spaceskip=\fontdimen2\font plus
\BIBentryALTinterwordstretchfactor\fontdimen3\font minus
  \fontdimen4\font\relax}
\providecommand{\BIBforeignlanguage}[2]{{%
\expandafter\ifx\csname l@#1\endcsname\relax
\typeout{** WARNING: IEEEtran.bst: No hyphenation pattern has been}%
\typeout{** loaded for the language `#1'. Using the pattern for}%
\typeout{** the default language instead.}%
\else
\language=\csname l@#1\endcsname
\fi
#2}}
\providecommand{\BIBdecl}{\relax}
\BIBdecl

\bibitem{5}
B.~Gurakan, O.~Ozel, J.~Yang, and S.~Ulukus, ``Energy cooperation in energy
  harvesting communications,'' \emph{IEEE Trans. Commun.}, vol.~61, no.~12, pp.
  4884--4898, Dec. 2013.

\bibitem{6}
K.~Tutuncuoglu and A.~Yener, ``Energy harvesting networks with energy
  cooperation: procrastinating policies,'' \emph{IEEE Trans. Commun.}, vol.~63,
  no.~11, pp. 4525--4538, Nov. 2015.

\bibitem{7}
A.~Biason and M.~Zorzi, ``Joint transmission and energy transfer policies for
  energy harvesting devices with finite batteries,'' \emph{IEEE J. Sel. Areas
  in Commun.}, vol.~33, no.~12, pp. 2626--2640, Dec. 2015.

\bibitem{1}
L.~R. Varshney, ``Transporting information and energy simultaneously,''
  \emph{Proc. IEEE Int. Symp. on Information Theory (ISIT)}, pp. 1612--1616,
  July 2008.

\bibitem{2}
P.~Grover and A.~Sahai, ``Shannon meets tesla: Wireless information and power
  transfer,'' \emph{Proc. IEEE Int. Symp. on Information Theory (ISIT)}, pp.
  2363--2367, June 2010.

\bibitem{3}
K.~Huang and E.~Larsson, ``Simultaneous information and power transfer for
  broadband wireless systems,'' \emph{IEEE Trans. Signal Processing}, vol.~61,
  no.~23, pp. 5972--5986, Sept. 2013.

\bibitem{4}
R.~Zhang and C.~K. Ho, ``{MIMO} broadcasting for simultaneous wireless
  information and power transfer,'' \emph{IEEE Trans. Wireless Commun.},
  vol.~12, no.~5, pp. 1989--2001, Mar. 2013.

\bibitem{8}
H.~Ju and R.~Zhang, ``Throughput maximization in wireless powered communication
  networks,'' \emph{IEEE Trans. Wireless Commun.}, vol.~13, no.~1, pp.
  418--428, Jan. 2014.

\bibitem{9}
------, ``User cooperation in wireless powered communication networks,''
  \emph{Proc. IEEE Global Communications Conference (GLOBECOM)}, pp.
  1430--1435, Dec. 2014.

\bibitem{10}
M.~A. Abd-Elmagid, T.~ElBatt, and K.~G. Seddik, ``Optimization of wireless
  powered communication networks with heterogeneous nodes,'' \emph{Proc. IEEE
  Global Communications Conference (GLOBECOM)}, Dec. 2015.

\bibitem{11}
A.~Biason and M.~Zorzi, ``Battery-powered devices in {WPCN}s,'' \emph{IEEE
  Trans. Commun.}, vol.~65, no.~1, pp. 216--229, Oct. 2016.

\bibitem{12}
M.~A. Abd-Elmagid, A.~Biason, T.~ElBatt, K.~G. Seddik, and M.~Zorzi, ``On
  optimal policies in full-duplex wireless powered communication networks,''
  \emph{Proc. Int. Symp. Modeling and Optimization in Mobile, Ad Hoc and
  Wireless Networks (WiOpt)}, pp. 243--249, May 2016.

\bibitem{13}
M.~A. Abd-Elmagid, T.~ElBatt, and K.~G. Seddik, ``A generalized optimization
  framework for wireless powered communication networks,''
  \emph{arXiv:1603.01115}, Mar. 2016.

\bibitem{14}
P.~D. Diamantoulakis, K.~N. Pappi, Z.~Ding, and G.~K. Karagiannidis, ``Wireless
  powered communications with non-orthogonal multiple access,''
  \emph{arXiv:1511.01291v2}, Feb. 2016.

\bibitem{15}
Y.~Saito, A.~Benjebbour, Y.~Kishiyama, and T.~Nakamura, ``System-level
  performance evaluation of downlink non-orthogonal multiple access ({NOMA}),''
  \emph{Proc. IEEE Int. Symp. on Personal, Indoor, and Mobile Radio
  Communications (PIMRC)}, pp. 611--615, Sept. 2013.

\bibitem{powercast}
``Powercast corporation, {TX}91501 user's manual \& {P}2110's datasheet.''

\bibitem{17}
M.~Avriel, \emph{Ed. Advances in Geometric Programming}, ser. Mathematical
  Concepts and Methods in Science and Engineering.\hskip 1em plus 0.5em minus
  0.4em\relax Plenum Press, 1980, vol.~21.

\bibitem{18}
M.~Chiang, \emph{Geometric programming for communication systems}, ser.
  Foundations and Trends in Communications and Information Theory, July 2005,
  vol.~2, no. 1--2.

\bibitem{20}
S.~Boyd and L.~Vandenberghe, \emph{Convex optimization}.\hskip 1em plus 0.5em
  minus 0.4em\relax Cambridge university press, 2004.

\bibitem{19}
B.~R. Marks and G.~P. Wright, ``A general inner approximation algorithm for
  nonconvex mathematical programs,'' \emph{Operations Research}, vol.~26,
  no.~4, pp. 681--683, 1978.

\bibitem{cvx}
M.~Grant and S.~Boyd, ``{CVX}: Matlab software for disciplined convex
  programming, version 2.1,'' \url{http://cvxr.com/cvx}, Mar. 2014.

\end{thebibliography}
    \bibliographystyle{IEEEtran}
\end{document}